\documentclass[12pt]{article}

\newcommand{\be}{\begin{eqnarray}}
\newcommand{\ee}{\end{eqnarray}}

\usepackage{amsfonts}

\def\Ro{{\mathbb R}}
\def\Tr {{\rm Tr}\,}

\def\psibar{\overline\psi}

\begin{document}

\title {A generalized quantum microcanonical ensemble}
\author{Jan Naudts and Erik Van der Straeten\\
{\small Departement Fysica, Universiteit Antwerpen,} \\
{\small Groenenborgerlaan 171, 2020 Antwerpen, Belgium}\\
{\small E-mail Jan.Naudts@ua.ac.be, Erik.VanderStraeten@ua.ac.be}
}

\maketitle

\begin {abstract}
We discuss a generalized quantum microcanonical ensemble.
It describes isolated systems that are not necessarily in an eigenstate of the Hamilton operator.
Statistical averages are obtained by a combination of a time average
and a maximum entropy argument to resolve the lack of knowledge about initial conditions.
As a result, statistical averages of linear observables coincide with values
obtained in the canonical ensemble. Non-canonical averages can be obtained by taking into account
conserved quantities which are non-linear functions of the microstate.
\end {abstract}

\section {Introduction}

In a recent paper, Goldstein et al \cite {GLTZ05} argue that the equilibrium state
of a finite quantum system should not be described by a density operator $\rho$
but by a probability distribution $f(\psi)$ of wavefunctions $\psi$ in
Hilbert space $\cal H$. The relation between both is then
\be
\rho=\int_{\cal S}{\rm d}\psi\,f(\psi)\,|\psi\rangle\,\langle\psi|.
\label {density}
\ee
Integrations over normalised wavefunctions may seem weird,
but have been used before in \cite {PDN93} and in papers quoted there.
Goldstein et al develop their arguments for a canonical ensemble at inverse temperature $\beta$.
For the microcanonical ensemble at energy $E$ they follow the old works of
Schr\"odinger and Bloch and define the equilibrium distribution as the uniform
distribution on a subset of wavefunctions which are linear combinations of
eigenfunctions of the Hamiltonian, with
corresponding eigenvalues in the range $[E-\epsilon,E+\epsilon]$. The parameter $\epsilon$ is
fixed and small.
However, there is no compelling reason why an isolated system should be in an eigenstate of
the Hamiltonian, or in a superposition of stationary states all with about the same energy.
On the contrary, time-dependent states are generic ---
their time average determines the equilibrium state. Such a time-dependent
state cannot be decomposed generically into eigenstates with energy in the range $[E-\epsilon,E+\epsilon]$.
Its wavefunction $\psi$ only satisfies the conditions
\be
\langle\psi|\,\psi\rangle=1
\qquad\hbox{ and }\qquad
\langle\psi|\,H\psi\rangle=E.
\label {manifold}
\ee
Hence, in a generalized microcanonical ensemble,
it is obvious to define the microcanonical density operator $\rho$ by means of
(\ref {density}), where the integration extends over all normalised wavefunctions $\psi$
satisfying (\ref {manifold}). 
This approach can be worked out explicitly, as has been done in a recent paper by Brody et al \cite {BHH05,BCH05}.
However, we believe that this is not yet the correct way of
defining the generalized quantum microcanonical ensemble. Indeed, the density operator $\rho$
obtained in this way has non-physical properties, even for simple $n$-level systems.
E.g., the heat capacity of the 2-level system does not depend on energy, contradicting
experimental evidence that heat capacity vanishes as the energy approaches the
ground state energy. The 4-level system exhibits a phase transition, which is totally unexpected.
In our opinion, this approach has to be modified in two aspects: 1) Due to the occurrence of non-commuting
observables replacing time averages by phase space averages is not straightforward;
2) The abundance of conserved quantities in quantum systems must be taken into account.

Recently, the interest in the microcanonical ensemble has increased because of the possibility of
microcanonical phase transitions, occurring even in finite systems \cite {GDH01}. The phenomenon
of negative specific heat was known in astrophysics. Its existence has been demonstrated rigorously in the
model of Hertel and Thirring \cite {HT71}.
When such an instability occurs then the usual argument of equivalence of ensembles becomes invalid. Hence,
one cannot anymore profit from the freedom of choice to use the most simple ensemble, but one is forced to do
microcanonical calculations. Up to now these are almost exclusively done using classical mechanics. However,
there are problems, such as the stability of metal clusters \cite {GFFGBCM01,RNGM03} or of atomic nuclei \cite {ZGXZ87},
that one would like to treat quantum mechanically. The standard definition of the quantum microcanonical
ensemble is useless for this purpose because it limits itself to (nearly) stationary states. By including
time-dependent states the behaviour of the system approaches that of its classical equivalent. This raises
the hope that a microcanonical phase transition, present in the classical system, shows up in the quantum system as well.

Conserved quantities are discussed in the next section. In Section 3 microcanonical averages are defined.
The classical ergodic theorem is used to replace time averages by integrations over wavefunctions.
In Section 4, the non-uniqueness of the equilibrium state is resolved by means of the maximum entropy principle.
As a result, the microcanonical averages coincide with canonical averages.
This takes away the need of proving the physical relevance of the present approach. Indeed,
the evidence that the canonical ensemble gives an adequate description of physical phenomena
is overwhelming. One should therefore expect only minor differences with the standard treatment of
quantum statistical mechanics. One of these is that the microcanonical ensemble is described
by a set of equiprobable wavefunctions, and not only by a density operator. A new feature is the possibility of
calculating statistical averages of nonlinear functions of the state of the system, while quantum
expectations of a quantum observable always depend linearly on the state of the system.

Sections 5 and 6 analyse the role of additional conserved quantities for the example of the harmonic oscillator.
Coherent states, modified with logarithmic corrections, are introduced.
The final section contains a preliminary evaluation of the present approach.

\section {Time averages and lack of ergodicity}

The ergodic theorem plays an important role in understanding the
microcanonical ensemble for classical (i.e.~non-quantum) systems. Even when the
ergodic hypothesis is not satisfied the statement remains that first averaging
over time and then over initial conditions in phase space gives the same result
as directly averaging over phase space alone. The only condition for
this to be true is that the probability distribution over phase space is time invariant.
In the quantum context the time average cannot be omitted and is an essential step in
obtaining a density matrix which commutes with the Hamiltonian.

Assume a Hamiltonian $H$ with energy eigenvalues $E_n$ and corresponding eigenstates $\psi_n$.
Let $\psi=\sum_n\lambda_n\psi_n$ be a wavefunction of the desired energy
\be
\langle\psi|H\psi\rangle=\sum_n|\lambda_n|^2E_n=E.
\label {energycond}
\ee
Its time evolution is given by
\be
\psi_t=\sum_n\lambda_ne^{-i\hbar^{-1}E_nt}\psi_n.
\ee
Then the time averaged expectation value of an operator $A$ is given by
\be
\lim_{T\rightarrow\infty}\frac 1T\int_0^T{\rm d}t\,\langle\psi_t|A\psi_t\rangle
=\Tr\rho_\psi A
\label {timavop}
\ee
with
\be
\rho_\psi=\sum_n|\lambda_n|^2 |\psi_n\rangle\langle\psi_n|.
\label {rhomicrocan}
\ee
With a strict definition of (generalised) microcanonical ensemble, any density operator
of the form (\ref {rhomicrocan}), with $\lambda_n$ satisfying $\sum_n|\lambda_n|^2=1$
and (\ref  {energycond}), is a candidate for describing the equilibrium distribution.
But this is clearly any density operator $\rho$ which is diagonal together with $H$
and which satisfies $\Tr\rho H=E$. This abundance of stationary states is a consequence
of the large number of conserved quantities. From a classical point of view this means that a quantum
system is almost always non-ergodic.

\section {Microcanonical averages}

An example of a non-linear function of the state of the system is
\be
f(\psi,\psibar)=\langle\psi|AB\psi\rangle-\langle\psi| A\psi\rangle\,\langle\psi| B\psi\rangle,
\label {quadrfun}
\ee
where $A$  and $B$ are quantum observables. The statistical average of such a function
is defined by
\be
\langle f\rangle_\psi=\lim_{T\rightarrow\infty}\frac 1T\int_0^T{\rm d}t\,f(\psi_t,\psibar_t),
\label {timav}
\ee
whenever this limit converges.
Note that $\psi=\sum_n\lambda_n\psi_n$ with $H\psi_n=E_n\psi_n$ implies that
$\psi_t=\sum_n\lambda_n\exp(-i\hbar^{-1}E_nt)\psi_n$.
Hence, by the classical ergodic theorem one has
\be
\langle f\rangle_\psi=\int{\rm d}\chi\,
f\left(\sum_n\lambda_ne^{i\chi_n}\psi_n, \sum_n\overline\lambda_ne^{-i\chi_n}\psi_n\right),
\label {ergodthm}
\ee
where
\be
\int{\rm d}\chi\equiv\prod_n\left(\frac 1{2\pi}\int_0^{2\pi}{\rm d}\chi_n\right).
\ee
The latter is an integration over wavefunctions, as in (\ref {density}). However,
the domain of integration is limited to a time-invariant subset of wavefunctions
generated by a single $\psi$.
Note that the ergodic hypothesis, needed for the validity of (\ref {ergodthm}),
is satisfied in the generic case. If it is not satisfied then, in line with the tradition of
classical statistical physics,
(\ref {ergodthm}) is taken as the definition of the microcanonical average rather than (\ref {timav}).

In the linear case is $f(\psi,\psibar)=\langle\psi|A\psi\rangle$.
Then (\ref {ergodthm}) becomes
\be
\langle f\rangle_\psi
&=&\int{\rm d}\chi\,\left\langle\sum_m\lambda_me^{i\chi_m}\psi_m|
A\sum_n\lambda_ne^{i\chi_n}\psi_n\right\rangle\cr
&=&\sum_n|\lambda_n|^2\langle\psi_n|A\psi_n\rangle\cr
&=&\Tr\rho_\psi A.
\ee
The average of the non-linear function (\ref {quadrfun}) becomes
\be
\langle f\rangle_\psi
&=&\Tr\rho_\psi AB-\int{\rm d}\chi\,
\left\langle\sum_m\lambda_me^{i\chi_m}\psi_m\bigg|
A\sum_n\lambda_ne^{i\chi_n}\psi_n\right\rangle\cr
& &\times
\left\langle\sum_{m'}\lambda_{m'}e^{i\chi_{m'}}\psi_{m'}\bigg|
B\sum_{n'}\lambda_{n'}e^{i\chi_{n'}}\psi_{n'}\right\rangle\cr
&=&\Tr\rho_\psi AB-(\Tr\rho_\psi A)(\Tr\rho_\psi B)\cr
& &
-\Tr\rho_\psi A\rho_\psi B
+\sum_n|\lambda_n|^2\langle\psi_n| A\psi_n\rangle\,\langle\psi_n| B\psi_n\rangle.
\ee
The last two terms are a consequence of the non-linearity.

\section {Maximum entropy}

The non-uniqueness of the microcanonical equilibrium state can be lifted in more than one way.
One approach would be to integrate (\ref {timavop}) over all normalised wavefunctions $\psi$
satisfying (\ref {energycond}). This can be worked out but leads to some strange results.
In particular, the resulting density operator $\rho$ depends non-analytically on $E$.
We do not pursue this approach.

The lack of information about the coefficients $\lambda_n$ can be dealt with by
means of the maximum entropy principle. Let $\mu_n=|\lambda_n|^2$.
Take entropy of the form
\be
S(\mu)=-k_{\rm B}\sum_n\mu_n\ln\mu_n,
\label {canent}
\ee
as usual.
After introduction of the Lagrange multipliers $\alpha$ and $\beta$, to take the
constraints $\sum_n\mu_n=1$ and $\sum_n\mu_nE_n=E$ into account, one obtains
\be
\mu_n=e^{-\alpha-\beta E_n}.
\label {canch}
\ee
But then the density operator $\rho_\psi$, defined by (\ref {rhomicrocan}),
coincides with the Boltzmann-Gibbs-von Neumann result of the canonical ensemble.

This choice of microcanonical wavefunctions has many advantages. First of all,
most microcanonical results coincide with the canonical ones. This eliminates the
need to convince the community that the microcanonical results are physically acceptable.
Further advantages follow from the use of the maximum entropy principle.
The canonical density operator is the most likely density operator satisfying the
energy requirement. It minimises the amount of information produced by the model.
But in addition, it guarantees thermodynamical stability. Indeed, the entropy (\ref {canent})
can be identified with thermodynamic entropy. Its derivative with respect to
energy $E$ is inverse temperature, and equals $k_B\beta$.

The choice (\ref {canch}) appears to be very satisfactory.
Nevertheless, some further refinement may be required, as will be
made clear in the next sections for the example of the harmonic oscillator.

\section {The harmonic oscillator}

We now start to develop an alternative approach based on the example of the harmonic oscillator.
The next section shows then how to reconcile this approach with the maximum entropy approach
discussed above. The Hamiltonian of the harmonic oscillator is
\be
H=\frac 1{2m}P^2+\frac 12m\omega^2Q^2
\ee
Its eigenvalues are $E_n=\hbar\omega(\frac 12+n)$, $n=0,1,\cdots$.

In the spirit of Boltzmann's entropy we look for a measure of the phase space extent
of the set of wavefunctions $\psi_t,t\in\Ro$. To this purpose, consider a semi-classical
phase portret obtained by measuring average position $Q$ and average momentum $P$.
Note that the exact solution of Heisenberg's equations of motion is known
\be
Q_t&=&Q\cos\omega t+\frac 1{m\omega}P\sin\omega t\\
P_t&=&-m\omega Q\sin\omega t+P\cos\omega t.
\ee
Hence, for any $\psi$, the averages $\langle Q\rangle_t=\langle\psi_t|Q\psi_t\rangle=\langle\psi|Q_t\psi\rangle$
and $\langle P\rangle_t=\langle\psi_t|P\psi_t\rangle=\langle\psi|P_t\psi\rangle$
describe ellipses in the $(\langle P\rangle,\langle Q\rangle)$-plane. The logarithm of the 'size'
of the ellipse can be taken as definition of (non-equilibrium) entropy.
It is well-known that the ellipse is maximal when $\psi$ is a coherent state wavefunction, i.e.
\be
\psi_z=e^{-|z|^2/2}\sum_{n=0}^\infty\frac 1{\sqrt {n!}}z^n\psi_n
\ee
for some complex $z$.
The maximal values, that are attained, equal
\be
\langle Q\rangle_{\rm max}&=&\lambda_Q|z|\qquad\hbox{ with }\lambda_Q=\sqrt {\frac {2\hbar}{m\omega}}\\
\langle P\rangle_{\rm max}&=&\lambda_P|z|\qquad\hbox{ with }\lambda_P=\sqrt {2\hbar m\omega}.
\ee
The energy condition for such a wavefunction reads
\be
E=\hbar\omega\left(\frac 12+|z|^2\right).
\ee

It is now obvious to claim that the coherent states are the equilibrium states of the microcanonical ensemble.
But this claim is incompatible with the maximum entropy argument of Section 4. Indeed, it implies
a Poisson distribution of the amplitudes $|\lambda_n|^2$ instead of an exponential distribution.
The way out of this conflict is the introduction of a microcanonical ensemble with two
conserved quantities.

\section {Additional conserved quantities}

Note that classical energy
\be
E_{\rm cl}(\psi)=\frac 1{2m}\langle\psi|P\psi\rangle^2+\frac 12m\omega^2\langle\psi|Q\psi\rangle^2
\ee
is a conserved quantity of the quantum harmonic oscillator.
Hence it is meaningful to study a microcanonical ensemble where both
$E_{\rm cl}(\psi)$ and quantum energy $E=\langle\psi|H\psi\rangle$ are fixed.
The equilibrium state of this ensemble is described by the wavefunction
$\psi=\sum_n\lambda_n\psi_n$ obtained by maximising
\be
-\sum_n|\lambda_n|^2\ln|\lambda_n|^2-\alpha\sum_n|\lambda_n|^2-\beta\sum_n|\lambda_n|^2E_n
-\gamma E_{\rm cl}(\psi).
\label {max}
\ee
Introduce creation and annihilation operators $a$ and $a^*$, as usual.
A short calculation shows
\be
E_{\rm cl}(\psi)
&=&\hbar\omega\langle\psi|a^*\psi\rangle\,\langle\psi|a\psi\rangle\cr
&=&\hbar\omega\left|\zeta\right|^2,
\label {afluct}
\ee
with
\be
\zeta=\sum_n\overline\lambda_{n+1}\lambda_n\sqrt{n+1}.
\ee
Variation of (\ref {max}) with respect to $|\lambda_n|^2$ then gives
\be
0&=&-\ln|\lambda_n|^2-1-\alpha-\beta E_n\cr
& &-\gamma\hbar\omega\Re\left[
\overline\zeta
\left(\frac {\overline\lambda_{n+1}}{\overline\lambda_n}\sqrt{n+1}+\frac {\lambda_{n-1}}{\lambda_n}\sqrt n\right).
\right]
\label {vareq}
\ee
This equation can be read as a recurrence relation for the coefficients $\lambda_n$.
Fixing $\lambda_0$, it determines a unique
normalised wavefunction, which is denoted $\psi(\beta,\gamma,\lambda_0)$.
Let us analyse the asymptotic behaviour of the $\lambda_n$. Put
\be
\lambda_n=\left(\frac 12\gamma \overline\zeta \hbar\omega\right)^n\frac {x_n}{c_n}
\label {asbeh}
\ee
with $x_n$ real and with $c_n=\prod_{p=0}^n(\sqrt{p}\ln\sqrt{p})$.
Then (\ref {vareq}) becomes
\be
& &x_n\left[2\ln c_n-2\ln x_n-n\ln \left(\frac 14\gamma^2 |\zeta|^2(\hbar\omega)^2\right)
-1-\alpha-\beta\hbar\omega\left(\frac 12+n\right)\right]\cr
&=&x_{n+1}\frac {|\zeta|^2(\gamma \hbar\omega)^2}{\ln(n+1)}
+x_{n-1}n\ln n.
\ee
From this expression follows that the $x_n$ are constant for large $n$ (note that $2\ln c_n$ increases as $n\ln n$).
Hence, (\ref {asbeh}) shows that the $\lambda_n$ behave asymptotically in the same manner
as the expansion coefficients of a coherent wavefunction, be it with logarithmic corrections
because $c_n\not=\sqrt {n!}$.

\section {Discussion}

We started with the observation that one cannot simply replace quantum time averages by
averages over some phase space of quantum mechanical microstates. Indeed, time averages
are essential to obtain a statistical description by a density operator which commutes
with the Hamiltonian of the system. But, restricting statistical mechanics to a formalism based
on nothing but time averages is not very satisfactory because of the intrinsic non-ergodicity
of quantum systems. The non-unicity may be resolved by means of the maximum entropy principle.
As a result, the microcanonical ensemble is described by the same density operator as the canonical
ensemble. This is not bad in itself because the results of the quantum canonical ensemble are
widely believed to agree with experimental reality. However, this cannot be the whole story.
We expect that microcanonical systems can be thermodynamically unstable, while canonically this can only
happen in the thermodynamic limit.

The microcanonical ensemble does not any longer coincide with the canonical ensemble when constraints
are introduced, which depend non-linearly on the microstate. We have tried out this idea on the
harmonic oscillator. Based on semi-classical arguments one expects that the coherent states are the
microcanonical equilibrium states of the harmonic oscillator. Hence the obvious question is wether
there exist conserved quantities such that coherent states are obtained by maximising entropy under
the constraint that the conserved quantities have given values. The quantum harmonic oscillator
has a conserved quantity which, for convenience, we have called the classical energy.
It controls whether the microstate is more quantum-like or more classical. It vanishes for eigenstates
and has its maximal value for coherent states.
For intermediate values, the microcanonical equilibrium states behave asymptotically in the same way
as coherent states.

The microcanonical ensemble describes closed systems. This means that the system plus
environment is in a product state. This state must however be prepared by the experimenter starting from a
situation where the system interacts with its environment. It has been shown recently that under rather
general conditions on system plus environment the reduced density operator of the system
is that of the canonical ensemble \cite {GLTZ05bis}. It is then no surprise that, after isolating the
system, it is still described by a density operator of the canonical ensemble. The only
correction to this view that we want to make is that one should take into account additional
conserved quantities, for example, those which control the degree of quantumness of the system.

\section *{Acknowledegments}
We thank Prof.  D.H.E. Gross for a clarifying discussion.

\section *{}


\begin {thebibliography}{99}
\raggedright\parskip 4pt

\bibitem {GLTZ05}
S. Goldstein, J. Lebowitz, R. Tumulka, N. Zanghi,
{\sl On the distribution of the wave function for systems in thermal equilibrium,}
to appear in J. Stat. Phys.,
arXiv:quant-ph/0309021v2 (2005).

\bibitem {PDN93} D.N. Page, {\sl Average entropy of a subsystem,} Phys. Rev. Lett. {\bf 71}(9), 1291-1294 (1993).

\bibitem {BHH05} D.C. Brody, D.W. Hook, L.P. Hughston,
{\sl Microcanonical distributions for quantum systems,}
arXiv:quant-ph/0506163.

\bibitem {BCH05} C.M. Bender, D.C. Brody, D.W. Hook,
{\sl Solvable model of quantum microcanonical states,}
J. Phys. A: Math. Gen. {\bf 38} L607-L613 (2005),
arXiv:quant-ph/0508004.

\bibitem {GDH01} D.H.E. Gross, {\sl Microcanonical Thermodynamics: Phase transitions in "small" systems,}
Lecture Notes in Physics 66 (World Scientific, 2001).

\bibitem {HT71} P. Hertel and W. Thirring, {\sl A soluble model for a system with negative specific heat,} Ann. Phys. 63, 520 (1971).

\bibitem {GFFGBCM01} F. Gobet, B. Farizon, M. Farizon, M. J. Gaillard, J. P. Buchet, M. Carr\'e, P. Scheier, T. D. M\"ark,
{\sl Direct Experimental Evidence for a Negative Heat Capacity in the Liquid-to-Gas Phase Transition in Hydrogen Cluster Ions:
Backbending of the Caloric Curve,} Phys. Rev. Lett. {\bf 89}, 183403 (2002).

\bibitem {RNGM03} J.A. Reyes-Nava, I. L. Garz\'on, K. Michaelian,
{\sl Negative heat capacity of sodium clusters, } Phys. Rev. B{\bf 87}, 203401 (2003).

\bibitem {ZGXZ87} Zhang Xiao-Ze, D. H. E. Gross, Xu Shu-Yan, Zheng Yu-Ming, 
{\sl On the decay of very hot nuclei (I). Canonical metropolis sampling of multifragmentation,}
Nucl. Phys. A{\bf 461}, 641-667 (1987);
{\sl On the decay of very hot nuclei (II). Microcanonical metropolis sampling of multifragmentation,}
Nucl. Phys. A{\bf 461}, 668-690 (1987).

\bibitem {GLTZ05bis} S. Goldstein, J.L. Lebowitz, R. Tumulka, N. Zanghi,
{\sl Canonical typicality,} Phys. Rev. Lett. {\bf 96}, 050403 (2006),
arXiv:cond-mat/0511091.

\end {thebibliography}
\end {document}